\documentclass[12pt]{iopart}

\usepackage{psfig}

\begin{document}

\newcommand{\suin}{\sum \hspace*{-5mm} \int\limits}

\letter{On the possibility of ``correlation cooling'' of ultracold neutral
plasmas}
\author{T Pohl, T Pattard and JM Rost}
\address{Max Planck Institute for the Physics of Complex Systems,
N\"othnitzer Stra{\ss}e 38, D-01187 Dresden, Germany}
\ead{tpohl@mpipks-dresden.mpg.de}

\begin{abstract}
Recent experiments with ultracold neutral plasmas show an intrinsic heating
effect based on the development of spatial correlations.  We investigate
whether this effect can be reversed, so that imposing strong spatial
correlations could in fact lead to cooling of the ions. We find that cooling
is indeed possible. It requires, however, a very precise preparation of the
initial state. Quantum mechanical zero-point motion sets a lower limit for
ion cooling.
\end{abstract}

\submitto{\jpb}

\maketitle
Experimentally, ultracold ($T \ll 1$ K) neutral plasmas could
be realized only recently by photoionizing a cloud of ultracold atoms
\cite{Kil99,Kul00,Kil01}. From a plasma physics perspective, they are very
appealing since they might provide laboratory realizations of so-called
strongly coupled plasmas, where the Coulomb interaction (inversely
proportional to the mean inter-particle spacing $a$)
dominates the random thermal motion (proportional to the temperature $T$).
As a consequence, interesting ordering phenomena such as Coulomb
crystallization can be observed. In order to be in this strong-coupling
regime, a system has to be either very dense (large Coulomb interaction) or
very cold (little thermal energy). The first scenario is realized in some
astrophysical contexts, which are difficult to access
experimentally. Hence, the possibility to realize a strongly coupled
plasma in the laboratory by going to ultralow temperatures offered
exciting prospects within the plasma physics community. In \cite{Kil99},
an ultracold neutral plasma has been created by photoionizing a gas of
laser-cooled Xe atoms. The extremely low temperatures of the Xe gas
suggested that the system was well within the strong-coupling regime, the
so-called {\sl Coulomb coupling parameter} $\Gamma \equiv e^2/(a k_{\rm B} T)$,
i.e.\ the ratio of potential to thermal energy, being of the order of $10$ for
the electrons and even up to $30000$ for the ions\footnote{The large mass
difference between electrons and ions implies a very long timescale for
energy transfer between the two subsystems. Hence, for practical purposes it
is justified to treat them as (almost) independent and assign different
temperatures to both subsystems.}. 

However, it was realized
quickly \cite{Kuz02,Rob02,Ger03} that despite the low temperature of the Xe
gas a strongly coupled plasma could not be created in this way.
Due to the fact that the neutral Xe atoms initially present
interact only very weakly, the plasma is created in a
completely uncorrelated non-equilibrium state. Hence, the subsequent conversion
of potential into kinetic energy rapidly heats both the
electron and ion subsystem, suppressing the development of substantial
correlations \cite{Kuz02,Mur01}.
This effect has been termed ``disorder-induced heating'' \cite{Ger03} or
``correlation-induced heating'' \cite{Rob04}, expressing the fact that it is
the development of spatial correlations during the
relaxation of the plasma towards an equilibrium state which leads to an
increase in temperature. Put another way, in the initial non-equilibrium state
the potential energy is higher than in the corresponding
equilibrium state of a plasma of the same average density. Consequently,
relaxation towards thermodynamic equilibrium will reduce the potential
energy in the system and increase the kinetic energy, i.e.\ the
temperature\footnote{In view of the fact that the initial state of the plasma
is far from equilibrium, it is not really justified to associate a
temperature with it, the properties of the system are not those of a state
in thermodynamical equilibrium with the same kinetic energy. However,
{\sl before} the photoionization occurs the atomic gas has a well-defined
temperature, and it is the temperature of these atoms that can be compared
to that of the ions after equilibration.}.
Additionally, the electrons are heated by three-body
recombination as well as continuum lowering. Suppression of electron heating
is thus much more difficult than simply reducing the correlation heating. At
present, there is a proposal of mixing the plasma with a gas of Rydberg atoms,
so that collisional ionization of these Rydberg atoms removes energy from
the electron component of the plasma \cite{Van04}.

For the creation of a
strongly coupled ionic component, different procedures have been
suggested to avoid these heating effects, namely
({\em i}) using fermionic atoms cooled below the Fermi temperature in the
initial state, so that the Fermi hole around each atom prevents the occurence
of small interatomic distances \cite{Mur01}; ({\em ii}) an intermediate step
of exciting atoms into high Rydberg states, so that the interatomic spacing is
at least twice the radius of the corresponding Rydberg state \cite{Ger03};
and ({\em iii}) the continuous laser-cooling of the plasma ions after their
initial creation, so that the correlation heating is counterbalanced by the
external cooling \cite{Kil03,Poh03}.  So far, none of these procedures has been
realized experimentally, however it has been shown
theoretically that continuous laser-cooling can indeed lead to a strongly
coupled plasma, connected with crystallization of the expanding system
\cite{Poh03}. In this case, the correlation heating effect is still present
and is offset by an additional external cooling mechanism. Proposals ({\em i})
and ({\em ii}), on the other hand, aim at suppressing the
heating directly by avoiding the small interparticle distances leading to a
large potential energy, i.e.\ by introducing (spatial) correlation in the
initial state. As has already been noted in \cite{Ger03}, this could also be
achieved using optical lattices to arrange the atoms.

At this point, the question arises naturally whether the correlation heating
could not only be avoided, but whether it can actually be reversed and
turned into cooling. As argued above, correlation heating describes the
fact that the uncorrelated ions created in the photoionization process have a
higher potential energy than in the thermodynamical equilibrium state. Hence,
equilibration will reduce the potential energy and increase the kinetic
energy of the system, making the ions hotter than the atoms in the initial
gas state. On the other hand, it is conceivable that the atomic gas might
be prepared in an ``overcorrelated'' state, where the potential energy is
lower than its equilibrium value. In this case, equilibration
must increase the potential energy on the cost of the kinetic energy, i.e.\ the
temperature must {\em decrease}.
This effect of ``correlation cooling'' has
been described before in \cite{Ger03b} and, based on a kinetic approach,
in more general terms in
\cite{Sem99}. In \cite{Ger03b}, the overcorrelated initial state was achieved
by suddenly decreasing the Debye screening length, i.e.\ the temperature of
the electrons, rather than by inducing spatial order. It was shown that in this
way, a reduction in temperature of the order of a few percent does indeed occur for the temperatures and densities considered.

In the present letter, we explore an alternative scheme for achieving
correlation cooling, namely the use of an optical lattice to induce spatial
correlations in the initial state as suggested in \cite{Ger03}. We describe,
both analytically and by molecular dynamics simulations,
the situation of a fully ordered initial state where the atoms have been
arranged in a perfect bcc-type lattice
structure using an optical lattice. In reality, their spatial distribution
will be broadened, classically due to their non-vanishing kinetic energy
corresponding to the initial-state temperature, quantum mechanically
corresponding to the eigenstates of the harmonic oscillator potential
generated by the optical lattice. This broadening is neglected in our
numerical simulations, however, its influence will be discussed in the
analytical considerations below.
For the sake of numerical simplicity, our initial state corresponds to a
sphere cut out from a bcc lattice rather than a more realistic cylindrical
shape, arguing that edge effects should be of minor importance if the number
of atoms in the system is large enough. The system is  propagated using the
hybrid molecular dynamics method described in \cite{Poh03}. Briefly, an
adiabatic approximation is employed for the electrons, and their distribution
is obtained from the steady-state King distribution \cite{Kin66}, widely used
in the study
of globular clusters. The ions are then propagated using the electronic
mean-field obtained from the steady-state distribution and by fully taking
into account the ion-ion interactions.
In the experiments under consideration, quasineutral plasmas with
$\Gamma_{\rm{e}}$
as low as $4\cdot10^{-3}$ \cite{Kul00} have been realized,
making Debye screening effects negligible,
while ion numbers $N_{\rm i}$ up to $10^8$ have been obtained \cite{Sim03}.
For our simulations,
such a large particle number would lead to a prohibitively
large numerical effort. However, the typical
timescale for the plasma expansion is $t_{\rm exp} \approx
\sqrt{\Gamma_{\rm e}} N_{\rm i}^{1/3} \omega_0^{-1}$,
where $\omega_0 = \sqrt{e^2 / M a^3}$ is the Einstein
frequency in the Wigner-Seitz model of a Coulomb solid and $M$ is the ion mass.
On the other hand, the relaxation of the ions takes place on a timescale of
$t_{\rm rel} \approx \omega_0^{-1}$.
Hence, the same macroscopic behavior can be obtained by decreasing $N_{\rm i}$
and increasing $\Gamma_{\rm{e}}$, keeping the ratio $t_{\rm exp} / t_{\rm rel}$
fixed. Therefore, for our simulations we chose $N_{\rm i} = 10^5$ and
an initial electron energy corresponding to $\Gamma_{\rm{e}} = 0.5$, while
the bare Coulomb potential was used for the ion-ion interaction.

\begin{figure}[tb]
\centerline{\psfig{figure=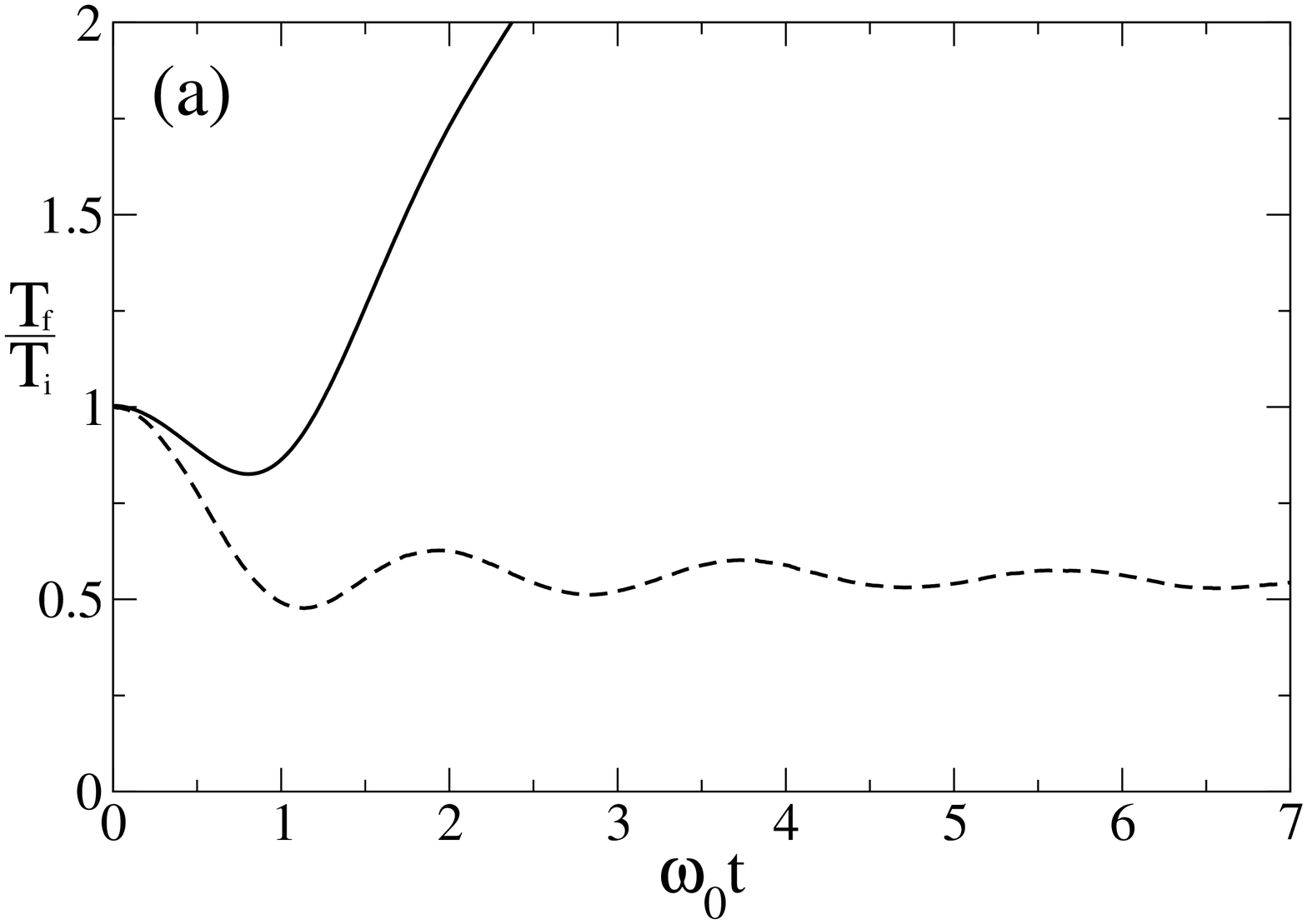,width=6.8cm} \hfill
\psfig{figure=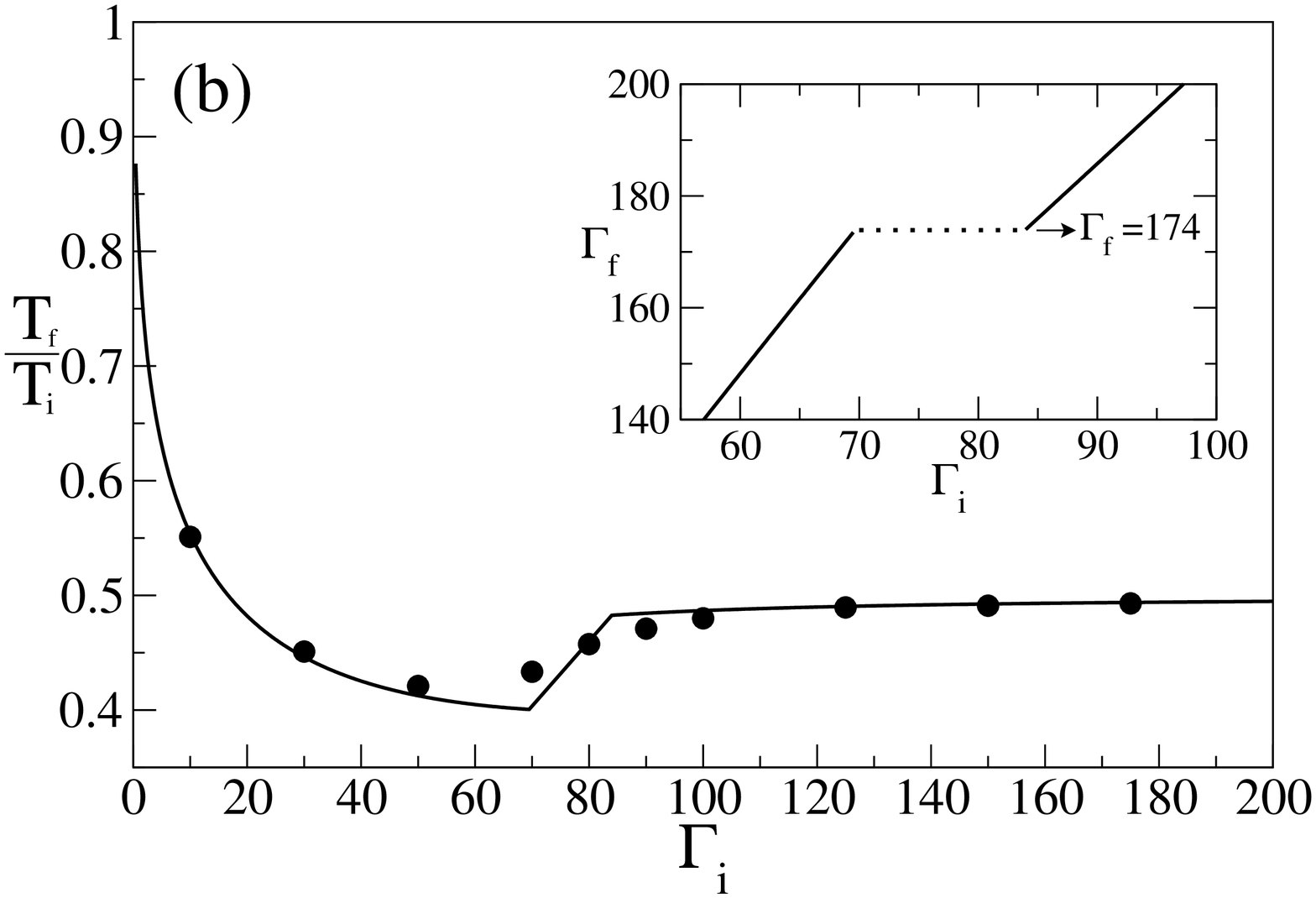,width=6.8cm}}
\caption{\label{fig1}
(a) Time evolution of the ionic kinetic energy, for the whole plasma cloud
(solid) and for the central region only (dashed).
(b) Ratio of final to initial temperature as a function of the density-scaled
inverse initial ion temperature $\Gamma_{\rm i}$, from the numerical simulations
(full
circles) and from equation (\ref{eq6}) (solid line). The inset shows the final
Coulomb coupling parameter.}
\end{figure}

Figure \ref{fig1}(a) shows the time evolution of the average ion kinetic energy in
units of the initial-state temperature for a typical realization. Since the
initial lattice configuration produces an effective oscillator potential
for each ion, the relaxation process is connected with transient oscillations
of the kinetic energy
which are damped out due to ion-ion collisions. Additionally, the average
kinetic energy increases at later times due to a radial ion acceleration caused by
the thermal electron pressure, which is proportional to the density
gradient. Therefore the
slow expansion of the unconfined plasma leads to a softening of the
initially sharp plasma edge and a modification of the velocity distribution
in the outer region of the plasma only. Since this distribution is close to a
Maxwellian in the central region of the plasma, the average kinetic energy is
directly related to the ionic temperature $T$, while it becomes more and more
dominated by the expansion energy towards the edge of the plasma.
In order to highlight these edge effects, we have calculated
the average ion kinetic energy from the velocity distribution in the inner
region only, taken to be a sphere with half of the initial plasma radius.
The resulting curve is shown as the dashed line in figure \ref{fig1}(a).
Clearly, the expansion hardly affects the kinetic energy in the inner region
of the plasma, leaving only the oscillations which
are damped out as the system approaches thermodynamic equilibrium. As
discussed above, the relaxation of the ions towards
equilibrium takes place on a timescale of $\omega_0^{-1}$, while the typical
timescale for the plasma expansion is $\sqrt{\Gamma_{\rm e}} N_{\rm i}^{1/3}
\omega_0^{-1}$.
Hence, a sufficiently large number of ions allows for an almost
unperturbed relaxation in the inner plasma region so that a temperature can
be defined in a meaningful way.
(For a strontium plasma with a typical density of
$5\cdot 10^9$cm$^-3$ used, e.g., in \cite{Sim03},
$\omega_0^{-1}\approx 0.2\:\mu$s, while
$t_{\rm exp}$ is about
$30$ times larger for $\Gamma_e=4\cdot 10^{-3}$ and $N_{i}=10^8$.)
Using the absorption imaging techniques
described in \cite{Sim03} it should be possible experimentally
to probe this inner region only. Therefore, we chose to define the final temperature
as given by the inner plasma region and to neglect the finite-size effects
associated with the boundary region of the plasma. The resulting $T_{\rm f}$
was estimated
from the center of the remaining oscillations and is shown in figure
\ref{fig1}(b) as a function of $\Gamma_{\rm i}$ by the full circles.
It should be noted that the value of $\Gamma_{\rm i}$ has no meaning
as a measure of nonideality of the initial plasma state, since the system may
be far from equilibrium. Therefore, in the present case it should simply be
viewed as a density-scaled inverse temperature rather than a Coulomb coupling
parameter. As can be seen in the figure, the ratio of final to initial
temperature reaches a
constant value of about 0.5 as the initial temperature goes to zero.
Hence, cooling can indeed be achieved on a moderate level of a
temperature reduction of about 50 percent.

For large particle numbers and if $\Gamma_{\rm e}$ is small enough such that
three-body recombination and Debye screening can be neglected, quasineutrality
leads to a homogeneous electron background in the central plasma region.
Thus, we can use the one-component plasma model to gain more insight into the
cooling process. This model has been studied extensively over the last
decades \cite{Ich82,Dub99}, which led to a detailed understanding of the
equation of state over a wide parameter range.
In the classical regime, the energy per particle of the final equilibrium state
is given by
\begin{equation} \label{eq1}
U_{\rm f}=\frac{3}{2}k_{\rm B} T_{\rm f} + W_{\rm f} \equiv (\frac{3}{2} +
w_{\rm f}) k_{\rm B} T_{\rm f} \;,
\end{equation}
where the excess energy $W$ accounts for nonideal effects due to the
inter-particle Coulomb interaction. For this quantity, several fit formulae
can be found in the literature, which have been obtained by Monte Carlo and
molecular dynamics methods \cite{Ich82,Dub99,Dew96}.
In the fluid phase we use the
interpolation formula from \cite{Cha98}
\begin{equation} \label{eq2}
w^{\rm fl}(\Gamma_{\rm f})=\Gamma_{\rm f}^{3/2}\left(\frac{A_1}{\sqrt{A_2+
\Gamma_{\rm f}}}+ \frac{A_3}{1+\Gamma_{\rm f}}\right) \; ,
\end{equation}
where $A_1=-0.9052$, $A_2=0.6322$ and $A_3=-\sqrt{3}/2-A_1/\sqrt{A_2}$.
Equation (\ref{eq2}) satisfies the known small-$\Gamma$ limit \cite{Abe59} and
accurately fits the high-$\Gamma$ values obtained by numerical calculations
\cite{Dew96}. In the solid phase ($\Gamma>174$), the excess energy can be
described by an expansion in powers of $1/\Gamma$ \cite{Dub99}
\begin{equation} \label{eq3}
w^{\rm s}(\Gamma_{\rm f})=-A_{\rm m}\Gamma_{\rm f}+\frac{3}{2}+\frac{a}{
\Gamma_{\rm f}} +\frac{b}{\Gamma_{\rm f}^2}+\frac{c}{\Gamma_{\rm f}^3}+
{\cal{O}}\left(\frac{1}{\Gamma_{\rm f}^4}\right) \; ,
\end{equation}
where $a=10.84$, $b=352.8$ and $c=1.794\cdot 10^5$.
Here, the first term corresponds to the Madelung energy of the lattice,
with $A_{\rm m}=0.895929$ for a bcc-type lattice. The second
term describes thermal contributions corresponding to an ideal gas of phonons
while the higher-order terms account for anharmonic corrections. Assuming an
unchanged ionic phase-space distribution function directly after the
photoionization, where the spatial distribution due to the finite temperature is
represented by Gaussian profiles centered on the sites of the bcc lattice,
the internal energy of the initial nonequilibrium state is given by
\begin{equation} \label{eq4}
U_{\rm i}= \frac{3}{2}k_{\rm B} T_{\rm i} + W_{\rm i} = \frac{3}{2}k_{\rm B}
T_{\rm i}-k_{\rm B} T_{\rm i} A_{\rm m}\Gamma_{\rm i}+k_{\rm B} T_{\rm i}
\frac{\Gamma_{\rm i}}{2} \left[\frac{\left<
r^2\right>}{a^2}-\frac{7}{8}B \left(\frac{\left<r^2\right>}{a^2}
\right)^2\right] \; ,
\end{equation}
where $B=\frac{1}{2}\sum_{i,j}\frac{a^5}{r^5_{ij}}$, with $B=0.876719$ for a
bcc lattice.
While the first two terms correspond to the ion kinetic energy and the Madelung
energy of a bcc lattice of point charges, the last two terms give the
leading-order corrections due to a broadened charge distribution at the
lattice sites expanded in terms
of the mean-squared ion displacement $\left<r^2\right>$. Assuming an initially
non-interacting atomic gas and harmonic trapping potentials we have
\begin{equation} \label{eq5}
\frac{\left<r^2\right>}{a^2}=\frac{3k_{\rm B}T_{\rm i}}{M\omega_{\rm L}^2a^2}=
3\frac{\omega_0^2}{\omega_{\rm L}^2}\Gamma_{\rm i}^{-1} \; ,
\end{equation}
where $\omega_{\rm L}$ is the oscillation frequency of the trapping
potentials.
The final temperature is now simply obtained from energy conservation
\begin{equation} \label{eq6}
\frac{T_{\rm f}}{T_{\rm i}}=\frac{3/2 + w_{\rm i}(\Gamma_{\rm i})}{3/2+w_{\rm f}
(\Gamma_{\rm f})} \; ,
\end{equation}
with $w = w^{\rm fl}$ for $\Gamma \le 174$ and $w = w^{\rm s}$ for $\Gamma >
174$. Figure \ref{fig1}(b) shows the ratio $T_{\rm f} / T_{\rm i}$ as a
function of $\Gamma_{\rm i}$ for $\left<r^2\right>/a^2=0$, i.e.\ for a
point-like ionic density
at the lattice sites. A comparison with our numerical data shows that equation
(\ref{eq6}) reproduces the numerical values for a finite-size plasma quite
well. The remaining discrepancies should be attributable to finite-size
effects \cite{Sch02,Has03}.
At small values of $\Gamma_{\rm i}$ the temperature ratio starts out from
unity and decreases with increasing $\Gamma_{\rm i}$, since the excess potential
energy of the final fluid state is naturally larger than that of the initial
lattice configuration. As $\Gamma_{\rm i}$ exceeds a value of $\Gamma_{\rm i}
\approx 70$
the temperature ratio rises linearly, connected with a solidification of the
plasma, where $\Gamma_{\rm f}=174$ stays constant (see inset). In this regime,
regions of fluid and solid phases coexist in the plasma. At
$\Gamma_{\rm i}\approx
84$ the system has crossed the melting point and the temperature ratio
approaches a value of $1/2$. This factor of $1/2$ originates from the
additional thermal
energy of $3k_{\rm B} T_{\rm f}/2$ arising from thermal lattice excitations.
However, the situation changes if a finite extension of the initial charge
distribution at the lattice sites is considered.
Neglecting terms of order $\left(\left<r^2\right>/a^2\right)^2$ in equation
(\ref{eq4}), one finds that for finite $\omega_{\rm L}$ the temperature
approaches a value of $(1+\omega_0^2/\omega_{\rm L}^2)/2$ if
$\left<r^2\right>\ll a^2$.

\begin{figure}[tb]
\centerline{\psfig{figure=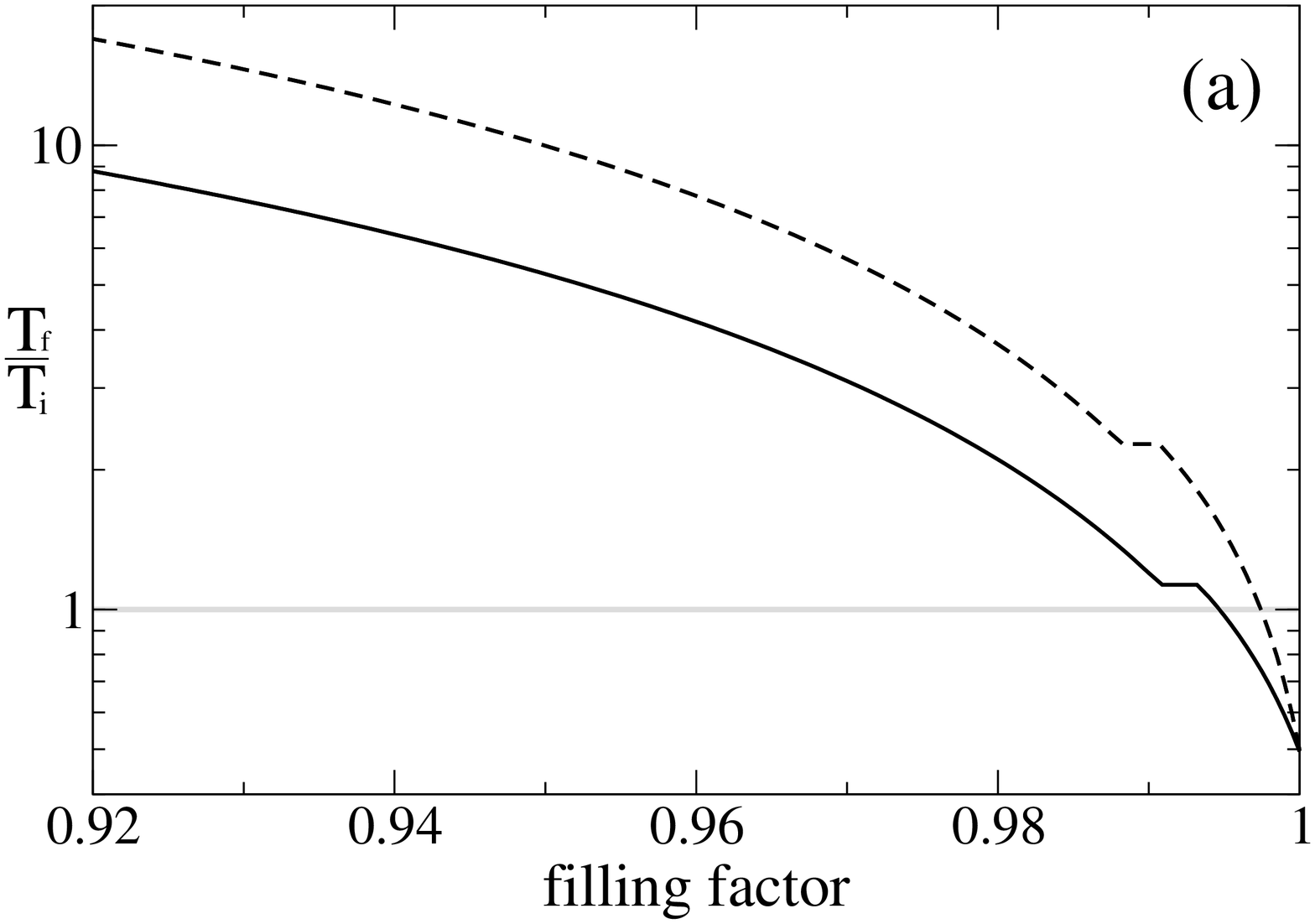,height=4.5cm} \hfill
\psfig{figure=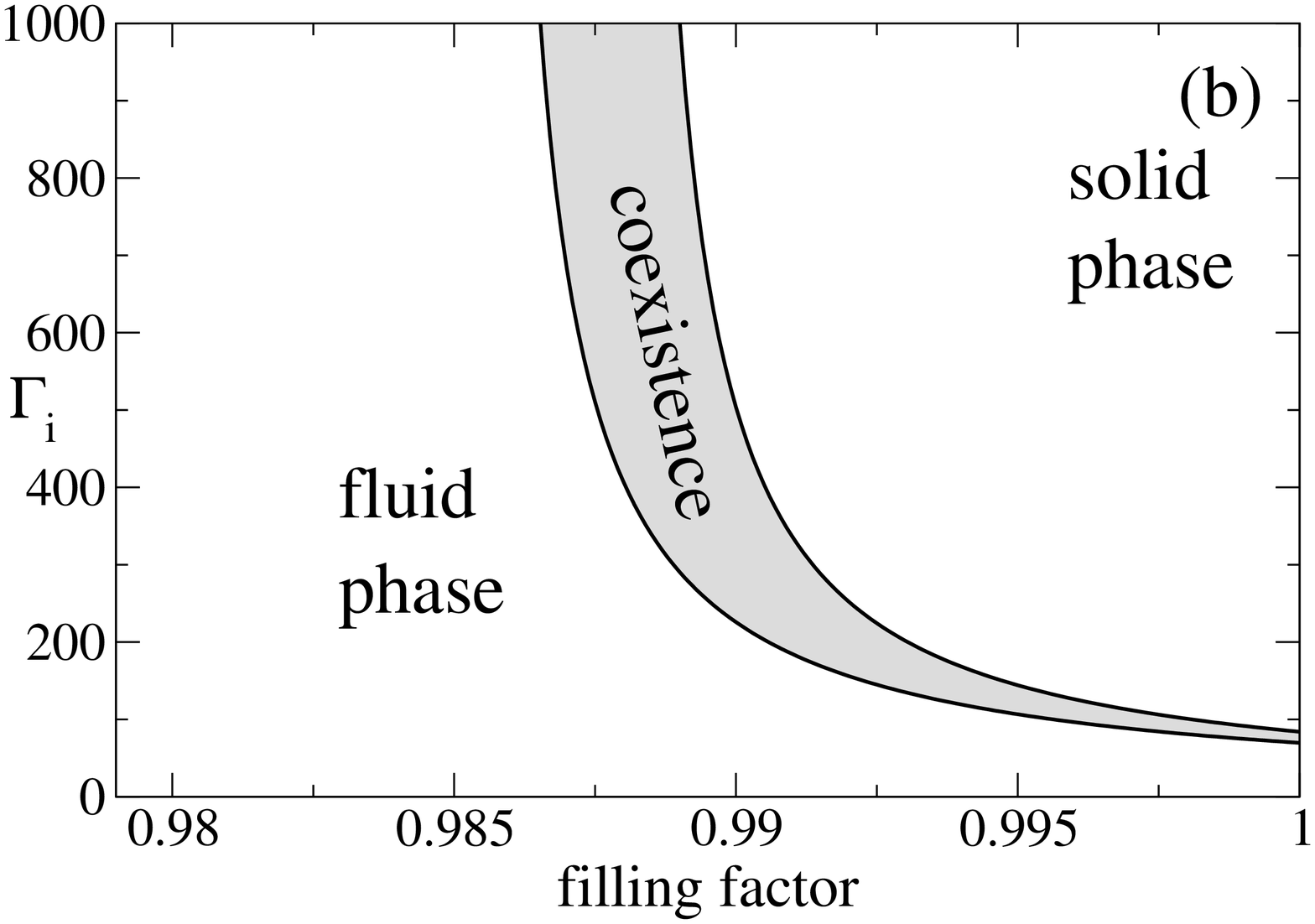,height=4.6cm}}
\caption{\label{fig2}
(a) Final temperature as a function of the lattice filling factor for
$\Gamma_{\rm i}=200$ (solid line) and $\Gamma_{\rm i}=400$ (dashed line).
The gray
horizontal line separates cooling from heating. (b) Phase diagram for the final
plasma state in the $f$-$\Gamma_{\rm i}$ plane.}
\end{figure}

Let us now turn to the question of the influence of possible lattice defects,
i.e.\ how perfect must the initial-state lattice be in order to observe
correlation cooling? Such defects are easily accounted for in the
previous consideration. Introducing a filling factor $f$, defined as the
probability that a given lattice site is occupied, simply reduces the average
charge at each lattice site. Therefore, the potential energy of the
initial state can be written as $f^2 w_{\rm i}(\Gamma_{\rm i}^{\star})
k_{\rm B} T_{\rm i}$, where $\Gamma_{\rm i}^{\star}$ corresponds to the
temperature scaled by the density
of the perfectly filled lattice, which is related to the actual value of
$\Gamma_{\rm i}$ by $\Gamma_{\rm i}^{\star}=f^{-1/3}\Gamma_{\rm i}$.
Substituting this in equation (\ref{eq4}) yields
\begin{equation} \label{eq7}
\frac{T_{\rm f}}{T_{\rm i}}=\frac{3/2+f^2w_{\rm i}(f^{-1/3}\Gamma_{\rm i})}{3/2
+w_{\rm f}(\Gamma_{\rm f})}
\end{equation}
as the generalization of equation (\ref{eq6}) to non-perfect filling.
In figure \ref{fig2}(a), we show the resulting final temperature as a function
of the filling factor for two different $\Gamma_{\rm i}$. A decreasing
filling enhances the initial potential energy (compared to that of a full
lattice with the same average {\em density}, {\em not} the same {\em lattice
constant}),
leading to a rapid increase of the final temperature. As a consequence,
cooling can be achieved for a nearly perfect
lattice only. Furthermore, for a fixed $f<1$, increasing $\Gamma_{\rm i}$
leads to
larger values for the temperature ratio. This heating also influences the
possibility of the creation of a crystallized plasma phase, since the melting
point of $\Gamma_{\rm f}=174$ is shifted to larger values of $\Gamma_{\rm i}$,
as shown in
the phase diagram figure \ref{fig2}(b). For values of $f<0.9855$, the system
always ends up in the fluid phase independent of $\Gamma_{\rm i}$.

\begin{figure}[tb]
\centerline{\psfig{figure=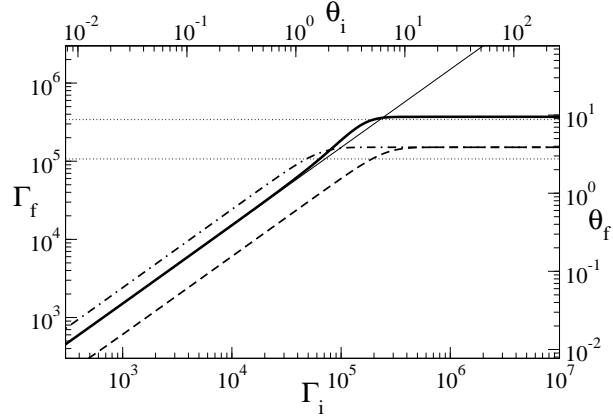,width=8cm}}
\caption{\label{fig3} 
Final scaled temperature as a function of the initial scaled temperature for
$r_{\rm s}=2 \cdot 10^9$ and for $\omega_{\rm L}/\omega_0=1$ (thick solid line),
$\omega_{\rm L}/\omega_0=2$ (dashed line) and $\omega_{\rm L}/\omega_0=1/2$
(dot-dashed line). For comparison, the result obtained with the classical
expressions for initial- and final-state energies is shown by the thin solid
line.
The horizontal dotted lines show the estimate
equation (\ref{eq11}) for the limiting temperature.}
\end{figure}

The above picture suggests that, for a perfect initial state, cooling is
always possible independent of the initial temperature. At low enough
temperatures, however, the classical treatment will break down, and quantum
mechanical effects will become important. As in the classical case the quantum
statistical internal energy of a Coulomb solid\footnote{As mentioned
previously, for typical experimental setups
as in \cite{Kil99}, $\Gamma_i$ is of the order of 10000, i.e.\ well within the
crystallized regime.} can be written as a sum of the Madelung energy, harmonic
contributions arising from linear lattice excitations and anharmonic
corrections. For the harmonic contributions to the excess energy we use an analytic approximation formula obtained recently
\cite{Bai01}, which reproduces the numerical results for harmonic Coulomb crystals very accurately.
The anharmonic corrections are taken from \cite{Iye93}, where they
were obtained by fitting the results of quantum Monte Carlo simulations. The
resulting expression is too lengthy to be reproduced here, but incorporates
both the known semiclassical high-temperature
limit as well as the quantum mechanical groundstate energy of a Coulomb
crystal,
\begin{equation} \label{eq8}
w_{\rm gs}= -A_{\rm m} \Gamma_{\rm f}+1.329 \theta-
\frac{0.365}{\sqrt{r_{\rm s}}}
\theta + {\cal{O}}\left(\frac{\theta}{r_{\rm s}^{3/2}}\right) \; .
\end{equation}
Here,
$\theta=(\hbar\omega_0)/(k_{\rm B} T)$ measures the importance of quantum
effects, i.e.\ zero-point ion oscillations, and $r_{\rm s}=aMe^2/\hbar^2$ is the Wigner-Seitz radius in units of the ionic Bohr
radius. 
As in the previous classical considerations, we apply a sudden approximation
assuming an unchanged density matrix directly after the
photoionization pulse. The initial energy of the plasma is thus found from
the expectation value of the new Hamiltonian using a density matrix
representing independent harmonic oscillators arranged on a bcc lattice.
The result is
\begin{equation} \label{eq9}
U_{\rm i}=\frac{3}{4}\hbar \omega_{\rm L}\coth{\left(\frac{\hbar
\omega_{\rm L}}{2k_{\rm B} T_{\rm i}}\right)}-k_{\rm B} T_{\rm i} A_{\rm m}
\Gamma_{\rm i} + k_{\rm B} T_{\rm i} \frac{\Gamma_{\rm i}}{2}\left[\frac{
\left<r^2\right>}{a^2}-\frac{7}{8}
B \left(\frac{\left<r^2\right>}{a^2}\right)^2\right]\; ,
\end{equation}
which exactly corresponds to the classical expression equation (\ref{eq4})
except that the kinetic energy $3 k_{\rm B} T_{\rm i} / 2$ is replaced by its
quantum mechanical counterpart. Moreover, the mean-squared displacement is
now given by
\begin{equation} \label{eq10}
\frac{\left<r^2\right>}{a^2}=\frac{3}{2} \nu \; {\rm{coth}}\left(\frac{\hbar
\omega_{\rm L}}{2k_{\rm B} T_{\rm i}}\right)
\end{equation}
rather than equation (\ref{eq5}), where $\nu=\omega_0/(\omega_{\rm L}
\sqrt{r_{\rm s}})$.
The final temperature is again obtained by equating the initial and final
energies of the plasma. Figure \ref{fig3} shows $\Gamma_{\rm f}$ as a function
of
$\Gamma_{\rm i}$ for $r_{\rm s}=2\cdot 10^{9}$. In contrast to the
classical case, here the final temperature cannot be decreased to arbitrarily
low values, rather it approaches a finite constant value as $T_{\rm i}
\rightarrow0$, since the initial charge distributions at the lattice sites
have a finite extension at $T_{\rm i}=0$ due to zero-point oscillations. Using
the low-temperature limit equation (\ref{eq8}) as an estimate for
the final energy and neglecting terms of order $\nu^2$ in the resulting
energy balance, this limiting temperature is found to be
\begin{equation} \label{eq11}
\frac{k_{\rm B} T_{\rm f}}{\hbar \omega_0}=\frac{1}{2}\frac{\omega_{\rm L}}{
\omega_0} +\frac{1}{2} \frac{\omega_0}{\omega_{\rm L}}-\Omega ,
\end{equation}
where $\Omega=0.886-0.243/\sqrt{r_{\rm s}}$. Since $\Omega < 0.886$
independent of $r_{\rm s}$,
the limiting temperature is always positive and takes on a minimum value of
\begin{equation} \label{eq12}
\left(\frac{k_{\rm B} T_{\rm f}}{\hbar \omega_0}\right)_{\rm{min}}=1-\Omega
\stackrel{r_{\rm s}\gg1}{\longrightarrow}0.114
\end{equation}
at an optimum ratio of $\omega_{\rm L}$ and $\omega_0$ of
\begin{equation} \label{eq13}
\left(\frac{\omega_{\rm L}}{\omega_0}\right)_{\rm{min}}=1 \;.
\end{equation}
Note, however, that the limiting temperature obtained numerically with the full
expression for $w_{\rm{f}}$ is slightly smaller since at finite temperature
the system will not be in the groundstate, hence its potential energy is
somewhat larger than that given by equation (\ref{eq8}). 
We may rewrite equation (\ref{eq12}) as
$\lambda_{\rm T}/a=\sqrt{2\pi/[\sqrt{r_{\rm
s}}(1-\Omega)]}\stackrel{r_{\rm s}
\gg1}{\longrightarrow}7.424r_{\rm s}^{-1/4}$, where
$\lambda_{\rm T}=h/\sqrt{2\pi M k_{\rm B} T_{\rm f}}$ is the thermal deBroglie
wavelength. This would suggest that it is possible to reach the degenerate
regime of $\lambda_{\rm T}/a\approx1$ if $r_{\rm s}$ can be decreased
sufficiently, i.e.\ to a value $r_{\rm s} \approx 3 \cdot 10^3$. Even for
hydrogen this would correspond to an unrealistically large density of
$4\cdot 10^{23}$cm$^{-3}$.
Note, however, that the value of $r_{\rm s}$ enters the ratio
$\lambda_{\rm T}
/ a$ with a power of $1/4$ only, so that values of $\lambda_{\rm T} / a =
1/10$ can already be reached if $r_{\rm s} \approx 3 \cdot 10^7$,
corresponding to a density of $4\cdot 10^{11}$cm$^{-3}$, which has already been
surpassed with cold Bose-condensed hydrogen atoms \cite{Fri98}. 
In this regime, the present considerations will not hold,
since effects of particle statistics will be important which have
not been included in the above expressions for the energies. However,
equation (\ref{eq12}) might serve as an estimate for the parameter values
necessary to reach the degenerate regime.

In summary, we have discussed the possibility of correlation cooling and of
the creation of a strongly coupled ultracold neutral plasma
by photoionization of a highly ordered state of cold atoms, achievable with
optical lattices. Classical considerations have shown that the
initial temperature can be reduced by a factor of two if the atomic sample is
noninteracting initially, offering the possibility to reach
ion temperatures which have not been achieved so far.
The quantum regime has
been investigated using existing expressions for the equilibrium internal
energy of the system. Here it was found that the final temperature approaches
a constant value which arises from zero-point oscillations in the lattice potentials and
depends on the density as well as the geometry of
the optical lattice.
An almost perfect lattice is
necessary to observe the discussed cooling effects, since even a small amount of lattice defects largely
increases the final temperature. However, it has been
demonstrated
that it is possible to create regular structures with
a well-defined number of atoms localized in the potential wells of an
optical lattice \cite{Gre02}.

Financial support from the DFG through grant RO1157/4 is gratefully
acknowledged.


\section*{References}
\begin{thebibliography}{20}
\bibitem{Kil99} Killian TC, Kulin S, Bergeson SD, Orozco LA, Orzel C and
Rolston SL 1999 \PRL {\bf 83} 4776

\bibitem{Kul00} Kulin S, Killian TC, Bergeson SD and Rolston SL 2000 \PRL
{\bf 85} 318

\bibitem{Kil01} Killian TC, Lim MJ, Kulin S, Dumke R, Bergeson SD and Rolston SL
2001 \PRL {\bf 86} 3759

\bibitem{Kuz02} Kuzmin SG and O'Neil TM 2002 \PRL {\bf 88} 065003

\bibitem{Rob02} Robicheaux F and Hanson JD 2002 \PRL {\bf 88} 055002

\bibitem{Ger03} Gericke DO and Murillo MS 2003 {\it Contrib.\ Plasma Phys.}
{\bf 43} 298

\bibitem{Mur01} Murillo MS 2001 \PRL {\bf 87} 115003

\bibitem{Rob04} Roberts JL, Fertig CD, Lim MJ and Rolston SL 2004
{\it Preprint} physics/0402041

\bibitem{Van04} Vanhaecke N, Comparat D, Tate DA and Pillet P 2004
{\it Preprint} quant-ph/0401045

\bibitem{Kil03} Killian TC, Ashoka VS, Gupta P, Laha S, Nagel SB, Simien CE,
Kulin S, Rolston SL and Bergeson SD 2003 \JPA {\bf 36} 6077

\bibitem{Poh03} Pohl T, Pattard T and Rost JM 2003 {\it Preprint}
physics/0311131

\bibitem{Ger03b} Gericke DO, Murillo MS, Semkat D, Bonitz M and Kremp D 2003
\JPA {\bf 36} 6087

\bibitem{Sem99} Semkat D, Kremp D and Bonitz M 1999 \PR E {\bf 59} 1557

\bibitem{Kin66} King IR 1966 {\it Astron.\ J.} {\bf 71} 64

\bibitem{Sim03} Simien CE, Chen YC, Gupta P, Laha S, Martinez YN, Mickelson
PG, Nagel SB and Killian TC 2003 {\it Preprint} physics/0310017

\bibitem{Ich82} Ichimaru S 1982 \RMP {\bf 54} 1017

\bibitem{Dub99} Dubin DHE and O'Neil TM 1999 \RMP {\bf 71} 87

\bibitem{Dew96} DeWitt H, Slattery W and Chabrier G 1996 {\it Physica} B
{\bf 228} 158

\bibitem{Cha98} Chabrier G and Potekhin AY 1998 \PR E {\bf 58} 4941

\bibitem{Abe59} Abe R 1959 {\it Prog.\ Theor.\ Phys.} {\bf 21} 475

\bibitem{Sch02} Schiffer JP 2002 \PRL {\bf 88} 205003

\bibitem{Has03} Hasse RW 2003 \jpb  {\bf 36} 1011

\bibitem{Bai01} Baiko DA, Potekhin AY and Yakovlev DG 2001 \PR E {\bf 64}
057402

\bibitem{Iye93} Iyetomi H, Ogata S and Ichimaru S 1993 \PR B {\bf 47} 11703

\bibitem{Fri98} Fried DG, Killian TC, Willmann L, Landhuis D,
Moss SC, Kleppner D and Greytak TJ 1998 \PRL {\bf 81}, 3811

\bibitem{Gre02} Greiner M, Mandel O, Esslinger T, H\"ansch TW and Bloch I 2002
{\it Nature} {\bf 415} 39

\end{thebibliography}
\end{document}